\documentclass[11pt,english]{article}
\usepackage[T1]{fontenc}
\usepackage[latin9]{inputenc}
\usepackage{geometry}
\usepackage{mathrsfs}
\usepackage{amsmath}
\usepackage{amssymb}
\usepackage{setspace}
\makeatletter
\newcommand{\lyxaddress}[1]{
	\par {\raggedright #1
	\vspace{1.4em}
	\noindent\par}
}

\usepackage[numbers,sort&compress]{natbib}
\usepackage{braket}
\usepackage[framemethod=tikz]{mdframed}

\newmdenv[
  hidealllines=false,
  backgroundcolor=pink!15,
  innerleftmargin=10pt,
  innerrightmargin=10pt,
  innertopmargin=5pt,
  innerbottommargin=10pt,
  leftmargin=5pt,
  rightmargin=5pt,
  splittopskip=20pt
]{shadedbox}

\let\originalleft\left
\let\originalright\right
\renewcommand{\left}{\mathopen{}\mathclose\bgroup\originalleft}
\renewcommand{\right}{\aftergroup\egroup\originalright}

\let\rket\ket
\let\rbra\bra
\let\rbraket\braket

\newcommand{\ketbra}[2]{\rket{#1}\!\rbra{#2}}
\let\rketbra\ketbra

\makeatother

\usepackage{babel}
\begin{document}
\title{Quantum Conditional Probabilities and New Measures of Quantum Information}
\author{Jacob A. Barandes$^{1,}$\thanks{barandes@physics.harvard.edu} $\ $and
David Kagan$^{2,}$\thanks{dkagan@umassd.edu}}
\maketitle

\lyxaddress{\begin{center}
\emph{\small{}$^{1}$Jefferson Physical Laboratory, Harvard University,
Cambridge, MA 02138}\\
\emph{\small{}$^{2}$Department of Physics, University of Massachusetts
Dartmouth, North Dartmouth, MA 02747}
\par\end{center}}
\begin{abstract}
We use a novel form of quantum conditional probability to define new
measures of quantum information in a dynamical context. We explore
relationships between our new quantities and standard measures of
quantum information, such as von Neumann entropy. These quantities
allow us to find new proofs of some standard results in quantum information
theory, such as the concavity of von Neumann entropy and Holevo's
theorem. The existence of an underlying probability distribution helps
shed light on the conceptual underpinnings of these results.
\end{abstract}

\global\long\def\ket#1{\rket{#1}}%
\global\long\def\bra#1{\rbra{#1}}%
\global\long\def\braket#1{\rbraket{#1}}%
\global\long\def\ketbra#1#2{\rketbra{#1}{#2}}%

\global\long\def\Uhat{\widehat{\mathcal{U}}}%
\global\long\def\Qhat{\widehat{\mathcal{Q}}}%
\global\long\def\Khat{\widehat{\mathcal{K}}}%
\global\long\def\Hhat{\widehat{\mathcal{H}}}%
\global\long\def\That{\widehat{\mathcal{T}}}%
\global\long\def\Ohat{\widehat{\mathcal{O}}}%

\newcommand{\rupdown}[2]{{{}^{#1}}_{#2}}
\global\long\def\updown#1#2{\rupdown{#1}{#2}}%

\newcommand{\rdownup}[2]{{{}_{#1}}^{#2}}
\global\long\def\downup#1#2{\rdownup{#1}{#2}}%

\makeatletter
\newcommand*{\mybig}[1]{{\hbox{$\scalebox{1.0}{#1}$}}}
\makeatother
\newcommand{\rbbrak}[1]{\raisebox{1.3pt}{\mybig[}#1\raisebox{1.3pt}{\mybig]}}

\global\long\def\bpar#1{\mybig(#1\mybig)}%
\global\long\def\bbrak#1{\rbbrak{#1}}%

\section{Introduction\label{sec:Introduction}}

Quantum information is primarily understood in terms of von Neumann
entropy and related quantities \cite{Wehrl1978,Nielsen2000}. Due
to inherently quantum phenomena such as entanglement, quantum information
measures\textemdash such as conditional von Neumann entropy and mutual
von Neumann information\textemdash lack well-defined underlying probability
distributions. Nevertheless, despite their own somewhat unclear conceptual
underpinnings, these quantities have proved useful for reframing and
clarifying aspects of quantum information. Many of the relationships
satisfied by classical information measures are mirrored by their
quantum analogues \cite{Umegaki1962,Wehrl1978,Nielsen2000}, sometimes
quite remarkably, as in the case of strong subadditivity \cite{LiebRuskai1973}.

In this paper, we define and study new forms of quantum information
that complement the standard quantities. The key ingredients in our
approach are conditional probability distributions, first studied
in \cite{Barandes2014,Barandes2020}, that provide an underlying picture
for the type of information being described. In particular, we are
able to provide a description of information flow in the context of
open quantum systems whose dynamical evolution is well-approximated
by linear, completely positive, trace-preserving (CPTP) maps, without
any explicit appeal to larger Hilbert spaces or ancillary systems.
We show that some standard results of quantum information theory emerge
quite naturally from our perspective.

Section \ref{sec:Background} provides some relevant background on
classical and quantum information. In Section \ref{sec:QCPandInformation},
we define new forms of quantum conditional entropy and quantum mutual
information in terms of quantum conditional probabilities, and briefly
describe a dynamical interpretation of these quantities. In Section
\ref{sec:EntropyGrowthDataProcessing}, we use the results of the
previous section to analyze processes under which there is growth
in entropy (in the sense of Shannon) and to provide new proofs of
the concavity of von Neumann entropy and quantum data processing.
We demonstrate that our quantum data-processing inequality provides
a natural interpretation of Holevo's theorem in a dynamical context,
showing that Holevo's $\chi$ acts as an upper bound on the amount
of information that can flow from a system's initial configuration
to a later one. In Section \ref{sec:Discussion}, we present a discussion
of various ways to generalize our constructions, including to an analysis
of the relationships between subsystems and the parent systems to
which they belong, and to more general decompositions of density matrices
than the ones that play a primary role in the paper. In Section \ref{sec:Connections},
we identify connections between the constructions in this paper and
previous work. We conclude in Section 7 with a brief summary of our
results and interesting open questions.

\section{Background\label{sec:Background}}

\subsection{Shannon Entropy, Density Matrices, and von Neumann Entropy}

Consider a classical random variable $X$ whose set of outcomes $\left\{ x\right\} _{x}$
occur according to a probability distribution $\left\{ p(x)\right\} _{x}$.
Using this data, we can compute expectation values, standard deviations,
and so on. Assuming a discrete set of outcomes, the average information
encoded in the probability distribution is given by its Shannon entropy:
\begin{equation}
H(X)\equiv-\sum_{x}p(x)\log p(x).\label{eq:ShannonEnt}
\end{equation}

In quantum theory, observables are a non-commutative generalization
of random variables, with their set of eigenvalues playing the role
of the set of possible outcomes. A given density matrix $\hat{\rho}$
generalizes the role of a probability distribution, allowing us to
compute statistical quantities such as the expectation value of an
observable $\hat{\mathcal{O}}$:
\begin{equation}
\langle\mathcal{O}\rangle=\textrm{Tr}[\hat{\rho}\hat{\mathcal{O}}].\label{eq:ExpVal}
\end{equation}

The simplest kind of density matrix corresponds to a pure state, and
can be expressed as a projection operator of the form $\ketbra{\Psi}{\Psi}$.
In this simple case, the formula (\ref{eq:ExpVal}) reduces to
\begin{equation}
\braket{\mathcal{O}}=\textrm{Tr}[\ketbra{\Psi}{\Psi}\hat{\mathcal{O}}]=\braket{\Psi\vert\hat{\mathcal{O}}\vert\Psi}.\label{eq:ExpVal-1}
\end{equation}
In general, a density matrix has infinitely many possible decompositions
over sets of projectors $\{\hat{\Pi}_{\alpha}\}_{\alpha}$,
\begin{equation}
\hat{\rho}=\sum_{\alpha}\lambda_{\alpha}\hat{\Pi}_{\alpha},\qquad\hat{\Pi}_{\alpha}=\ketbra{\phi_{\alpha}}{\phi_{\alpha}},
\end{equation}
where the set $\left\{ \lambda_{\alpha}\right\} _{\alpha}$ consists
of non-negative real numbers that sum to unity, and where $\left\{ \ket{\phi_{\alpha}}\right\} _{\alpha}$
is not necessarily an orthonormal set of states. Each such decomposition
has a corresponding Shannon entropy:
\begin{equation}
H\left(\left\{ \lambda_{\alpha}\right\} \right)=-\sum_{\alpha}\lambda_{\alpha}\log\lambda_{\alpha}.
\end{equation}
The decomposition that \emph{minimizes} \cite{Jaynes1957} the Shannon
entropy consists of the eigenvalues and corresponding eigenprojectors
of $\hat{\rho}$,
\begin{equation}
\hat{\rho}=\sum_{i}p_{i}\hat{P}_{i},\qquad\hat{P}_{i}=\ketbra{\Psi_{i}}{\Psi_{i}},
\end{equation}
where $\left\{ \ket{\Psi_{i}}\right\} _{i}$ is the set of eigenstates
of $\hat{\rho}$. The von Neumann entropy of a density matrix $\hat{\rho}$
is this minimal Shannon entropy of $\hat{\rho}$,
\begin{equation}
S(\hat{\rho})\equiv-\textrm{Tr}[\hat{\rho}\log\hat{\rho}]=-\sum_{i}p_{i}\log p_{i},
\end{equation}
and therefore represents the minimum amount of average information
that can be encoded in a system described by $\hat{\rho}$.

\subsection{Classical Conditional Entropy and its Quantum Counterpart}

Classically, the conditional entropy of a random variable $Y$ given
another random variable $X$ is defined in terms of a conditional
probability distribution $p(y\vert x)$ that describes correlations
between possible outcomes of the two random variables $Y$ and $X$.
Specifically, the conditional entropy of a random variable $Y$ given
that $X$ takes the value $x$ is defined to be
\begin{equation}
H(Y\vert x)\equiv-\sum_{y}p(y\vert x)\log[p(y\vert x)].
\end{equation}
The full conditional entropy is then
\begin{equation}
H(Y\vert X)\equiv\sum_{x}H(Y\vert x)p(x)=-\sum_{x,y}p(y\vert x)p(x)\log[p(y\vert x)],\label{eq:ClassicalConditionalEntropy}
\end{equation}
which can be thought of as the average information encoded in $Y$
given a particular outcome of $X$, averaged over all the possible
outcomes of $X$.

Conditional entropies satisfy the identity
\begin{equation}
H(Y\vert X)=H(Y,X)-H(X),\label{eq:ClassicalConditionalEntropyIdentity}
\end{equation}
where $H(Y,X)$ is the Shannon entropy of the joint distribution in
$X$ and $Y$. The identity (\ref{eq:ClassicalConditionalEntropyIdentity})
captures the intuition that the conditional entropy measures the information
about $Y$ encoded in its correlations with $X$ in excess of information
encoded in $X$ alone.

In the quantum case, the pair of random variables $X$ and $Y$ are
replaced by a bipartite quantum system $AB$, with a corresponding
density matrix $\hat{\rho}_{AB}$. The standard definition of conditional
von Neumann entropy adopts the form of the classical relation (\ref{eq:ClassicalConditionalEntropyIdentity}),
with $S(\hat{\rho}_{AB})$ in place of the classical joint entropy
and $S(\hat{\rho}_{B})$ substituted for $H(X)$, where $\hat{\rho}_{B}$
is the reduced density matrix for subsystem $B$, as defined by the
partial trace over subsystem $A$. That is, the conditional von Neumann
entropy is given by
\begin{equation}
S(A\vert B)\equiv S(\hat{\rho}_{AB})-S(\hat{\rho}_{B}),\qquad\hat{\rho}_{B}=\textrm{Tr}_{A}[\hat{\rho}_{AB}].\label{eq:StandardQCE}
\end{equation}
Unlike classical conditional entropy, conditional von Neumann entropy
defined by (\ref{eq:StandardQCE}) lacks an underlying probability
distribution, as can be seen from the fact that $S(A\vert B)$ can
be negative \cite{Wehrl1978} when subsystems $A$ and $B$ are entangled.
In \cite{Cerf1997}, the authors introduce a conditional amplitude
operator $\hat{\rho}_{A\vert B}$ as one possible generalization of
a conditional probability distribution, but the operator is not a
density matrix, and thus lacks a clear interpretation itself. Operational
approaches are quite fruitful (see \cite{Horodecki2007} for example),
but they do not always clarify the conceptual underpinnings of such
quantities.

\section{Quantum Conditional Probabilities and Information\label{sec:QCPandInformation}}

\subsection{Quantum Conditional Probabilities\label{subsec:Quantum-Conditional-Probability}}

The type of information measures studied in this paper are built from
quantum conditional probabilities first explored in the context of
the minimal modal interpretation of quantum theory \cite{Barandes2014,Barandes2020}.
While the quantities we discuss here require nothing beyond standard
quantum theory for their formulation, we adopt the language of the
minimal modal interpretation, as it provides a useful way to describe
what follows.

To start, imagine that at a given time, a quantum system is described
by an `objective' density matrix $\hat{\rho}_{Q}$\textemdash objective
in the sense that it is empirically optimal among all possible density
matrices that an external observer could assign to the system.\footnote{Specifically, by an objective density matrix, we mean a density matrix
whose mixedness arises entirely from entanglement to other systems
and is therefore a solely improper mixture. In particular, we do not
include any classical uncertainty. For a system not entangled with
its environment, the objective density matrix is a rank-one projector
representing a pure state. For a system entangled with its environment,
the von Neumann entropy for the objective density matrix is precisely
equal to the entanglement entropy. A physically realistic observer
cannot improve on a system's objective density matrix without physically
affecting the system by introducing new forms of entanglement.} Now suppose that from the initial time to a later time, the density
matrix evolves from $\hat{\rho}_{Q}$ to a final density matrix $\hat{\rho}_{R}$
according to a linear CPTP map $\mathcal{E}_{R\leftarrow Q}$: 
\begin{equation}
\hat{\rho}_{R}=\mathcal{E}_{R\leftarrow Q}\left\{ \hat{\rho}_{Q}\right\} .\label{eq:EvoDM}
\end{equation}
The initial and final density matrices have respective spectral decompositions
\begin{align}
\hat{\rho}_{Q} & =\sum_{q}p_{q}\hat{P}_{q},\qquad\hat{P}_{q}=\ketbra{\Psi_{q}}{\Psi_{q}},\label{eq:InitialObjectiveDM}\\
\hat{\rho}_{R} & =\sum_{r}p{}_{r}\hat{P}{}_{r},\qquad\hat{P}{}_{r}=\ketbra{\Psi{}_{r}}{\Psi{}_{r}}.\label{eq:FinalObjectiveDM}
\end{align}

According to the minimal modal interpretation, every quantum system
has an actual underlying state corresponding to one of the eigenstates
of the system's density matrix, but that actual underlying state is
hidden from external observers unless the system's density matrix
is a projector. In our present example, the system's actual underlying
state evolves from being one of the eigenstates of $\hat{\rho}_{Q}$
to being one of the eigenstates of $\hat{\rho}_{R}$. Collectively,
the eigenstates of $\hat{\rho}_{Q}$ represent the initial possible
underlying states of the system, and the eigenstates of $\hat{\rho}_{R}$
represent the final possible underlying states.

The evolution of the possible underlying states of the system is defined
stochastically in terms of quantum conditional probabilities. For
example, the probability that the system's later state is $\ket{\Psi_{r}}$
given that it was initially $\ket{\Psi_{q}}$ is defined to be 
\begin{equation}
p_{\mathcal{E}}(r\vert q)\equiv\textrm{Tr}\big[\hat{P}{}_{r}\mathcal{E}_{R\leftarrow Q}\{\hat{P}_{q}\}\big]=\braket{\Psi_{r}\vert\mathcal{E}_{R\leftarrow Q}\{\hat{P}_{q}\}\vert\Psi{}_{r}}.\label{eq:QCP}
\end{equation}
Note that throughout this paper, lower-case index labels $q,q',\ldots$
and $r,r',\ldots$ on states correspond respectively to upper-case
system configuration labels $Q$ and $R$. We adopt analogous conventions
for other system configuration labels.

Regardless of the interpretation of quantum theory, the quantities
defined by (\ref{eq:QCP}) exhibit almost all of the standard properties
of conditional probabilities. In particular, they are non-negative
real numbers that sum to unity and satisfy the law of total probability,
\begin{equation}
p_{r}=\sum_{q}p_{\mathcal{E}}(r\vert q)p_{q}.\label{eq:TotalProbability}
\end{equation}
To see this, observe that
\begin{equation}
p_{r}=\textrm{Tr}\big[\hat{P}{}_{r}\hat{\rho}_{R}\big].
\end{equation}
Substituting (\ref{eq:EvoDM}) for $\hat{\rho}_{R}$ yields
\begin{equation}
p_{r}=\textrm{Tr}\big[\hat{P}{}_{r}\mathcal{E}_{R\leftarrow Q}\{\hat{\rho}_{Q}\}\big].
\end{equation}
We now substitute the decomposition (\ref{eq:InitialObjectiveDM})
of $\hat{\rho}_{Q}$ and use the linearity of $\mathcal{E}_{R\leftarrow Q}$
to rewrite the expression as
\begin{equation}
p_{r}=\sum_{q}\textrm{Tr}\big[\hat{P}{}_{r}\mathcal{E}_{R\leftarrow Q}\{\hat{P}_{q}\}\big]p_{q},\label{eq:TotalProbDerv}
\end{equation}
allowing us to arrive at (\ref{eq:TotalProbability}) by identifying
the trace in (\ref{eq:TotalProbDerv}) as the quantum conditional
probabilty (\ref{eq:QCP}).

The quantum conditional probabilities $p_{\mathcal{E}}(r\vert q)$
can be associated with a formal density matrix
\begin{equation}
\hat{\rho}_{R\vert q}^{\mathcal{E}}\equiv\sum_{r}p_{\mathcal{E}}(r\vert q)\hat{P}_{r},\label{eq:ConditionalDM}
\end{equation}
which satisfies
\begin{equation}
\hat{\rho}_{R}=\sum_{q}p_{q}\hat{\rho}_{R\vert q}^{\mathcal{E}},\label{eq:TotalDM}
\end{equation}
due to (\ref{eq:TotalProbability}).

A crucial difference between classical and quantum conditional probabilities
is that the latter fail to satisfy Bayes' theorem:
\begin{equation}
p_{\mathcal{E}}(r\vert q)p_{q}\neq p_{\mathcal{E}}(q\vert r)p_{r}.\label{eq:NoBayes}
\end{equation}
The failure of Bayes' theorem reflects the non-commutativity of quantum
observables, and therefore the inability to define a symmetric joint
probability distribution. From a dynamical perspective, Bayes' theorem
fails due to the generic irreversibility of $\mathcal{E}_{R\leftarrow Q}$,
as is evident from the case in which $\mathcal{E}_{R\leftarrow Q}$
represents a projective measurement.\footnote{The paper by Schack, Brun, and Caves \cite{Schack2001} is a prominent
example of work that does indeed derive a quantum version of Bayes'
rule. However, these sorts of results rely on taking a large number
of copies of a system's Hilbert space to represent a large ensemble
of identical systems. The conditional probabilities we define in (\ref{eq:QCP})
differ in essential ways from these earlier constructions, as is apparent
from the fact that our conditional probabilities involve only a single
instance of a system's Hilbert space. Thus, the failure of Bayes'
theorem is compatible with these prior results.}

In general, linear CPTP evolution of an eigenprojector of the initial
density matrix yields a nontrivial density matrix defined by
\begin{equation}
\hat{\rho}_{q}^{R}\equiv\mathcal{E}_{R\leftarrow Q}\big\{\hat{P}_{q}\big\}.\label{eq:DMRq}
\end{equation}
Introducing a new label $r_{q}$ to distinguish the eigenprojectors
$\{\hat{P}_{r_{q}}\}_{r_{q}}$ of this density matrix, we can write
down its spectral decomposition:

\begin{equation}
\hat{\rho}_{q}^{R}=\sum_{r_{q}}p_{\mathcal{E}}(r_{q}\vert q)\hat{P}_{r_{q}}.\label{eq:rhoRqDecomp}
\end{equation}
Note that for each fixed value of $q$, the basis of eigenprojectors
$\{\hat{P}_{r_{q}}\}_{r_{q}}$ can be different, and will generically
differ from $\{\hat{P}_{r}\}_{r}$.

Nevertheless, the set of these density matrices must combine to yield
$\hat{\rho}_{R}$,
\begin{equation}
\hat{\rho}_{R}=\sum_{q}p_{q}\hat{\rho}_{q}^{R},\label{eq:SumDM}
\end{equation}
as a consequence of (\ref{eq:DMRq}).

The relations (\ref{eq:TotalDM}) and (\ref{eq:SumDM}) suggest that
$\hat{\rho}_{R\vert q}^{\mathcal{E}}$ and $\hat{\rho}_{q}^{R}$ are
themselves related. To see how, notice that the quantum conditional
probabilities $p_{\mathcal{E}}(r\vert q)$ can be expressed as
\begin{align}
p_{\mathcal{E}}(r\vert q) & =\textrm{Tr}\big[\hat{P}_{r}\hat{\rho}_{q}^{R}\big]\nonumber \\
 & =\sum_{r_{q}}p_{\mathcal{E}}(r_{q}\vert q)\textrm{Tr}\big[\hat{P}_{r}\hat{P}_{r_{q}}\big],\label{eq:BornProbTran1}
\end{align}
where in passing from the first to the second line we have used the
decomposition (\ref{eq:rhoRqDecomp}). The quantity inside the trace
has the form of a Born probability,
\begin{equation}
\beta(r\vert r_{q})\equiv\textrm{Tr}\big[\hat{P}_{r}\hat{P}_{r_{q}}\big]=\vert\braket{\Psi_{r}\vert\Psi_{r_{q}}}\vert^{2},\label{eq:BornProb}
\end{equation}
and therefore (\ref{eq:BornProbTran1}) takes the form of a law of
total probability,
\begin{equation}
p_{\mathcal{E}}(r\vert q)=\sum_{r_{q}}\beta(r\vert r_{q})p_{\mathcal{E}}(r_{q}\vert q).\label{eq:BornProbTransition}
\end{equation}

Substituting the relation (\ref{eq:BornProbTransition}) into the
definition (\ref{eq:ConditionalDM}) yields
\begin{align}
\hat{\rho}_{R\vert q}^{\mathcal{E}} & =\sum_{r,r_{q}}\beta(r\vert r_{q})p_{\mathcal{E}}(r_{q}\vert q)\hat{P}_{r}\nonumber \\
 & =\sum_{r,r_{q}}p_{\mathcal{E}}(r_{q}\vert q)\hat{P}_{r}\hat{P}_{r_{q}}\hat{P}_{r}\nonumber \\
 & =\sum_{r}\hat{P}_{r}\Big(\sum_{r_{q}}p_{\mathcal{E}}(r_{q}\vert q)\hat{P}_{r_{q}}\Big)\hat{P}_{r},
\end{align}
where in passing to the second line we have used
\begin{equation}
\hat{P}_{r}\hat{P}_{r_{q}}\hat{P}_{r}=\beta(r\vert r_{q})\hat{P}_{r}.
\end{equation}
We thus arrive at the relation

\begin{equation}
\hat{\rho}_{R\vert q}^{\mathcal{E}}=\sum_{r,r_{q}}\hat{P}_{r}\hat{\rho}_{q}^{R}\hat{P}_{r}.\label{eq:condRhoFromProjectiveMeasurement}
\end{equation}
Note that 
\begin{equation}
S(\hat{\rho}_{R\vert q}^{\mathcal{E}})\geq S(\hat{\rho}_{q}^{R}),\label{eq:CPTPMoreRandom}
\end{equation}
which follows from the double stochasticity of the Born probability
distribution $\beta(r\vert r_{q})$.\footnote{We discuss doubly stochastic probability distributions in the appendix,
providing an explicit proof of a generalization of (\ref{eq:CPTPMoreRandom}).}

So far, our description of the quantum conditional probabilities (\ref{eq:QCP})
has been dynamical, with $\mathcal{E}_{R\leftarrow Q}$ thought of
as an evolution map. However, the same ideas can be applied to the
quantum relationships between systems and their subsystems by noting
that partial traces are an example of a linear CPTP map. We provide
a more detailed sketch of these ideas in Section \ref{sec:Discussion}.
In what follows, we will continue to focus on the dynamical picture,
in which a single system evolves according to $\mathcal{E}_{R\leftarrow Q}$.

\subsection{New Measures of Quantum Information}

Combining the quantum conditional probabilities of (\ref{eq:QCP})
with Shannon's entropy formula yields a new type of quantum conditional
entropy. Using the initial and final density matrices defined in (\ref{eq:InitialObjectiveDM})
and (\ref{eq:FinalObjectiveDM}), respectively, we let
\begin{equation}
J_{\mathcal{E}}(R\vert q)\equiv-\sum_{r}p_{\mathcal{E}}(r\vert q)\log[p_{\mathcal{E}}(r\vert q)]=S(\hat{\rho}_{R\vert q}^{\mathcal{E}})\label{eq:QCPShannonInfo}
\end{equation}
be the quantum conditional entropy of our system given that the system's
initial underlying state corresponded to the eigenstate $\ket{\Psi_{q}}$
of $\hat{\rho}_{Q}$. We will argue that we can interpret this quantity
as the entropy added to the system during its evolution given the
initial underlying state of the system. The full quantum conditional
entropy is the average over all possible initial eigenstates of $\hat{\rho}_{Q}$:
\begin{equation}
J_{\mathcal{E}}(R\vert Q)\equiv\sum_{q}J_{\mathcal{E}}(R\vert q)p_{q}=-\sum_{q,r}p_{\mathcal{E}}(r\vert q)p_{q}\log[p_{\mathcal{E}}(r\vert q)].\label{eq:NewQCE}
\end{equation}

We also define a new type of quantum mutual information:
\begin{equation}
I_{\mathcal{E}}\left(R:Q\right)\equiv\sum_{q,r}p_{\mathcal{E}}(r\vert q)p_{q}\log\left[\frac{p_{\mathcal{E}}(r\vert q)}{p_{r}}\right].\label{eq:QCausalMutual}
\end{equation}
The relation
\begin{equation}
I_{\mathcal{E}}\left(R:Q\right)=S(\hat{\rho}_{R})-J_{\mathcal{E}}(R\vert Q)\label{eq:QCausalMutualAndQCE}
\end{equation}
follows directly from the definitions of quantum conditional entropy
(\ref{eq:NewQCE}) and quantum mutual information (\ref{eq:QCausalMutual}),
mirroring the classical identity
\begin{equation}
I(Y:X)=H(Y)-H(Y\vert X).
\end{equation}
In a dynamical context, mutual information can be thought of as measuring
the information that is shared between the initial and final system
configurations.

The new forms of quantum conditional entropy and quantum mutual information
defined in (\ref{eq:QCPShannonInfo}) and (\ref{eq:QCausalMutual}),
respectively, are distinct from the traditional quantities found in
the literature. As discussed in Section \ref{sec:Background}, the
traditional conditional von Neumann entropy $S(A\vert B)$ in equation
(\ref{eq:StandardQCE}) is not defined in terms of an underlying probability
distribution. The traditional von Neumann mutual information $I^{VN}(A:B)$
shared by subsystems $A$ and $B$ is defined as
\begin{equation}
I^{VN}(A:B)\equiv S(\hat{\rho}_{A})-S(A\vert B).\label{eq:TraditionalQMI}
\end{equation}
Once again, there need not be any underlying probability distribution
in these traditional definitions.

We will show that the new information measures developed in this paper
satisfy inequalities that are analogous to those satisfied by (\ref{eq:StandardQCE})
and (\ref{eq:TraditionalQMI}). However, the existence of underlying
quantum conditional probabilities (\ref{eq:QCP}) provides conceptually
clearer interpretations of the sort of information measured by these
new quantities.

\subsubsection{Evolution from a Pure State}

To illustrate the interpretations of the quantities (\ref{eq:QCPShannonInfo})
and (\ref{eq:QCausalMutual}), we examine two special cases. To start,
consider a system that is initially in a known pure state $\ket{\Psi}$.
Suppose that it evolves according to a linear CPTP map $\mathcal{E}$,
so that we lose track of its initially pure state:
\begin{equation}
\hat{\rho}_{R}=\mathcal{E}\{\hat{P}_{\Psi}\},\qquad\hat{P}_{\Psi}=\ketbra{\Psi}{\Psi}.
\end{equation}
In this situation, we have conditional probabilities
\begin{equation}
p_{\mathcal{E}}(r\vert\Psi)=p_{r},
\end{equation}
and hence we have the quantum conditional entropy
\begin{equation}
J_{\mathcal{E}}(R\vert Q)=J_{\mathcal{E}}(R\vert\Psi)=-\sum_{r}p_{r}\log p_{r}=S(\hat{\rho}_{R}).
\end{equation}
In words, the increase in the system's entropy arises solely from
the evolution of the system. We can also characterize this statement
in terms of the mutual information, which vanishes,

\begin{equation}
I_{\mathcal{E}}(R;\Psi)=S(\hat{\rho}_{R})-J_{\mathcal{E}}(R\vert\Psi)=0,
\end{equation}
thereby showing that no information is carried over from the system's
initial state to its final configuration.

This linear CPTP map can be thought of as modeling a process in which
the system becomes more entangled with its surrounding environment.\footnote{This interpretation assumes that the map is faithful to the underlying
physics, rather than capturing measurement or modeling errors.} From this perspective, the quantum conditional entropy measures the
growth of entanglement between a system and its environment.

\subsubsection{Unitary Evolution}

Now consider a system whose initial and final density matrices are
$\hat{\rho}_{Q}$ and $\hat{\rho}_{R}$, as expressed in (\ref{eq:InitialObjectiveDM})
and (\ref{eq:FinalObjectiveDM}), respectively. Suppose that the evolution
is unitary, so that for some unitary operator $\hat{U}$, we have
\begin{equation}
\hat{\rho}_{R}=\mathcal{U}_{R\leftarrow Q}\left\{ \hat{\rho}_{Q}\right\} =\hat{U}\hat{\rho}_{Q}\hat{U}^{\dagger},\qquad\hat{U}\hat{U}^{\dagger}=\hat{U}^{\dagger}\hat{U}=\mathbb{I},
\end{equation}
where $\mathbb{I}$ is the identity. Under such evolution, the eigenvalues
of $\hat{\rho}_{Q}$ are unchanged and the eigenstates rotate into
the set of eigenstates of $\hat{\rho}_{R}$,
\begin{equation}
\hat{P}_{q}^{R}=\hat{U}\hat{P}_{q}^{Q}\hat{U}^{\dagger},
\end{equation}
where the upper label emphasizes that the evolution carries us from
the initial configuration $Q$ to the final configuration $R$. In
this situation, the conditional probabilities (\ref{eq:QCP}) are
trivial,
\begin{equation}
p_{\mathcal{U}}(r\vert q)=\textrm{Tr}\big[\hat{P}_{r}^{R}\hat{U}\hat{P}_{q}^{Q}\hat{U}^{\dagger}\big]=\textrm{Tr}\big[\hat{P}_{r}^{R}\hat{P}_{q}^{R}\big]=\delta_{rq}.
\end{equation}
The quantum conditional entropy of this process is therefore zero
and the quantum mutual information is equal to the von Neumann entropy
of the system, showing that the uncertainty in the state of the system
before the evolution is the sole source of uncertainty in the state
afterward.

\subsection{Some Identities and Inequalities}

Due to the existence of an underlying probability distribution, the
quantum conditional entropy (\ref{eq:QCPShannonInfo}) and mutual
information (\ref{eq:QCausalMutual}) satisfy various relationships
familiar from classical information theory.
\begin{itemize}
\item Conditional entropy and mutual information are always non-negative:
\begin{equation}
J_{\mathcal{E}}(R\vert Q)\geq0,\qquad I_{\mathcal{E}}(R:Q)\geq0.\label{eq:InformationPositivity}
\end{equation}
\item A system's mutual information cannot be greater than the system's
initial entropy:
\begin{equation}
I_{\mathcal{E}}(R:Q)\leq S(\hat{\rho}_{Q}).\label{eq:MutualLessThanSinitial}
\end{equation}
\item A system's conditional entropy cannot be greater than the system's
final entropy:
\begin{equation}
J_{\mathcal{E}}(R\vert Q)\leq S(\hat{\rho}_{R}).\label{eq:ConditioningReducesEntropy}
\end{equation}
\end{itemize}
The inequalities (\ref{eq:InformationPositivity}), (\ref{eq:MutualLessThanSinitial}),
and (\ref{eq:ConditioningReducesEntropy}) can be proved following
similar steps to those from classical information theory. We provide
details in the appendix.

\section{Entropy Growth and Data Processing\label{sec:EntropyGrowthDataProcessing}}

\subsection{Unital Evolution and Projective Measurement}

A unital linear CPTP map satisfies
\begin{equation}
\mathcal{E}_{R\leftarrow Q}\left\{ \mathbb{I}\right\} =\mathbb{I}.
\end{equation}
The conditional probabilities (\ref{eq:QCP}) for a unital linear
CPTP map are doubly stochastic:
\begin{align}
\sum_{q}p_{\mathcal{E}}(r\vert q) & =\textrm{Tr}\Big[\hat{P}_{r}\mathcal{E}_{R\leftarrow Q}\Big\{\sum_{q}\hat{P}_{q}\Big\}\Big]\nonumber \\
 & =\textrm{Tr}\left[\hat{P}_{r}\mathcal{E}_{R\leftarrow Q}\left\{ \mathbb{I}\right\} \right]\nonumber \\
 & =\textrm{Tr}\left[\hat{P}_{r}\right]\nonumber \\
 & =1.\label{eq:DoublyStochastic}
\end{align}
Thus, if the evolution of a system is unital linear CPTP, then the
von Neumann entropy grows,
\begin{equation}
S(\hat{\rho}_{Q})\leq S(\hat{\rho}_{R}),\label{eq:RandomizingIncreasesEntropy}
\end{equation}
which follows from the law of total probability (\ref{eq:TotalProbability})
relating $p_{r}$ and $p_{q}$ and the double-stochasticity of $p_{\mathcal{E}}(r\vert q)$
in this case, as proved in the appendix.

A projective measurement without post-selection is an example of a
unital process. Suppose that we measure an observable with eigenstates
$\left\{ \ket{\Psi_{m}}\right\} _{m}$. If we isolate the measurement
device and refrain from learning the outcome, then the post-measurement
density matrix is well-approximated by
\begin{equation}
\text{\ensuremath{\hat{\rho}_{M}}}=\mathcal{M}\left\{ \hat{\rho}_{Q}\right\} =\sum_{m}\hat{P}_{m}\hat{\rho}_{Q}\hat{P}_{m},\label{eq:ProjectiveMeasurement}
\end{equation}
which is clearly unital. As a result, we see that measurements without
post-selection increase the entropy of a system.

\subsection{Concavity of von Neumann Entropy}

The quantities described earlier allow us to demonstrate certain standard
properties of quantum information. Consider the concavity of von Neumann
entropy,
\begin{equation}
\sum_{i}p_{i}S(\hat{\rho}_{i})\leq S(\hat{\rho}),\qquad\hat{\rho}=\sum_{i}p_{i}\hat{\rho}_{i},\label{eq:ConcavityvNE}
\end{equation}
where $\hat{\rho}$ is an arbitrary density matrix, and the set of
pairs $\left\{ (p_{i},\hat{\rho}_{i})\right\} _{i}$ is any collection
of non-negative weights and density matrices that form a decomposition
of $\hat{\rho}$ with the weights summing to unity. Note that the
number of elements in the set can exceed the dimension of the Hilbert
space.

To prove (\ref{eq:ConcavityvNE}), we let $\hat{\rho}=\hat{\rho}_{R}$.
Given a decomposition into a set of weights and density matrices $\left\{ (p_{i},\hat{\rho}_{i})\right\} _{i}$
we can define a linear CPTP map $\mathcal{E}$ and a density matrix
$\hat{\rho}_{Q}$ such that $\hat{\rho}_{R}=\mathcal{E}\{\hat{\rho}_{Q}\}$
such that the elements of the decomposition arise from $\mathcal{E}$
applied to the eigen-decomposition of $\hat{\rho}_{Q}$, with the
identification of the $i$ and $q$ indices.\footnote{Note that we implicitly allow $\mathcal{E}$ to involve a partial
trace operation so that the Hilbert space dimension associated with
the final density matrix $\hat{\rho}_{R}$ can be smaller than that
of $\hat{\rho}_{Q}$.} From the relations (\ref{eq:TotalDM}), (\ref{eq:SumDM}), and (\ref{eq:condRhoFromProjectiveMeasurement}),
we have,
\begin{equation}
\hat{\rho}_{R}=\sum_{q}p_{q}\hat{\rho}_{R\vert q}=\sum_{q}p_{q}\hat{\rho}_{q}^{R},
\end{equation}
with
\begin{equation}
\hat{\rho}_{R\vert q}=\sum_{r}\hat{P}_{r}\hat{\rho}_{q}^{R}\hat{P}_{r}.
\end{equation}
Note that we have simplified the notation by suppressing some labels.

The quantum conditional entropy (\ref{eq:NewQCE}) can be expressed
as the sum
\[
J_{\mathcal{E}}(R\vert Q)=\sum_{q}p_{q}S(\hat{\rho}_{R\vert q}).
\]
Thus,
\[
\sum_{q}p_{q}S(\hat{\rho}_{q}^{R})\leq\sum_{q}p_{q}S(\hat{\rho}_{R\vert q})\leq S(\hat{\rho}_{R}),
\]
where the first inequality follows from (\ref{eq:CPTPMoreRandom}),
while the second is the inequality (\ref{eq:ConditioningReducesEntropy}),
demonstrating the concavity of von Neumann entropy.

\subsection{Quantum Markovianity and Data Processing\label{subsec:Quantum-Markovianity-and}}

Consider a system that evolves from $\hat{\rho}_{Q}$ to $\hat{\rho}_{R}$,
and then to $\hat{\rho}_{S}$, as described by the linear CPTP maps
$\mathcal{E}_{R\leftarrow Q}$ and $\mathcal{E}_{S\leftarrow R}$,
so that we have
\begin{equation}
\hat{\rho}_{R}=\mathcal{E}_{R\leftarrow Q}\{\hat{\rho}_{Q}\},\qquad\hat{\rho}_{S}=\mathcal{E}_{S\leftarrow R}\{\hat{\rho}_{R}\}=\mathcal{E}_{S\leftarrow R}\circ\mathcal{E}_{R\leftarrow Q}\{\hat{\rho}_{Q}\}=\mathcal{E}_{S\leftarrow Q}\{\hat{\rho}_{Q}\}.
\end{equation}
Observe that
\begin{equation}
\hat{\rho}_{R}=\mathcal{E}_{R\leftarrow Q}\Big\{\sum_{q}p_{q}\hat{P}_{q}\Big\}=\sum_{q}p_{q}\mathcal{E}_{R\leftarrow Q}\{\hat{P}_{q}\},
\end{equation}
with corresponding conditional probabilities
\begin{equation}
p(r\vert q)=\textrm{Tr}\big[\hat{P}_{r}\mathcal{E}_{R\leftarrow Q}\big\{\hat{P}_{q}\big\}\big],
\end{equation}
where we suppress the map label as the mapping will be clear from
the state indices.

Similarly, we have
\begin{align}
\hat{\rho}_{S} & =\mathcal{E}_{S\leftarrow R}\Big\{\sum_{r}p_{r}\hat{P}_{r}\Big\}=\sum_{r}p_{r}\mathcal{E}_{S\leftarrow R}\big\{\hat{P}_{r}\big\},\qquad p(s\vert r)=\textrm{Tr}\big[\hat{P}_{s}\mathcal{E}_{S\leftarrow R}\big\{\hat{P}_{r}\big\}\big],
\end{align}
as well as
\begin{equation}
\hat{\rho}_{S}=\mathcal{E}_{S\leftarrow Q}\Big\{\sum_{q}p_{q}\hat{P}_{q}\Big\}=\sum_{r}p_{q}\mathcal{E}_{S\leftarrow Q}\big\{\hat{P}_{q}\big\},\qquad p(s\vert q)=\textrm{Tr}\big[\hat{P}_{s}\mathcal{E}_{S\leftarrow Q}\big\{\hat{P}_{q}\big\}\big].
\end{equation}

There are some subtle constraints required for the consistency of
these processes. Using the law of total probability and (\ref{eq:BornProbTransition}),
we have
\begin{align}
p_{s} & =\sum_{r}p(s\vert r)p_{r}\nonumber \\
 & =\sum_{r,q}p(s\vert r)p(r\vert q)p_{q}.\nonumber \\
 & =\sum_{r,q,r_{q}}p(s\vert r)\beta(r\vert r_{q})p(r_{q}\vert q)p_{q}.\label{eq:ps1}
\end{align}
Similarly, we have
\begin{equation}
p_{s}=\sum_{q}p(s\vert q)p_{q}.
\end{equation}
However, recall from (\ref{eq:DMRq}) that
\begin{equation}
\mathcal{E}_{R\leftarrow Q}\big\{\hat{P}_{q}\big\}=\sum_{r_{q}}p(r_{q}\vert q)\hat{P}_{r_{q}}.
\end{equation}
So expanding out the definition of $p(s\vert q)$ and using $\mathcal{E}_{S\leftarrow Q}=\mathcal{E}_{S\leftarrow R}\circ\mathcal{E}_{R\leftarrow Q}$
gives
\begin{align}
p_{s} & =\sum_{q}\textrm{Tr}\big[\hat{P}_{s}\mathcal{E}_{S\leftarrow R}\big\{\mathcal{E}_{R\leftarrow Q}\big\{\hat{P}_{q}\big\}\big\}\big]p_{q}\nonumber \\
 & =\sum_{q}\textrm{Tr}\Big[\hat{P}_{s}\mathcal{E}_{S\leftarrow R}\Big\{\sum_{r_{q}}p(r_{q}\vert q)\hat{P}_{r_{q}}\Big\}\Big]p_{q}\nonumber \\
 & =\sum_{q}\sum_{r_{q}}\textrm{Tr}\big[\hat{P}_{s}\mathcal{E}_{S\leftarrow R}\big\{\hat{P}_{r_{q}}\big\}\big]p(r_{q}\vert q)p_{q}\nonumber \\
 & =\sum_{q}\sum_{r_{q}}p(s\vert r_{q})p(r_{q}\vert q)p_{q}.\label{eq:ps2}
\end{align}
Comparing (\ref{eq:ps1}) and (\ref{eq:ps2}), we find that a natural-looking
consistency condition to impose would be
\begin{equation}
p(s\vert r_{q})=\sum_{r}p(s\vert r)\beta(r\vert r_{q}).\label{eq:consistent}
\end{equation}
We therefore restrict our maps $\mathcal{E}_{R\leftarrow Q}$ and
$\mathcal{E}_{S\leftarrow R}$ to those satisfying (\ref{eq:consistent}).
The existence of such maps can be demonstrated by expanding out the
definitions of the conditional probabilities in (\ref{eq:consistent})
on both sides. On the right-hand side we have
\begin{align}
\sum_{r}p(s\vert r)\beta(r\vert r_{q}) & =\sum_{r}\textrm{Tr}\big[\hat{P}_{s}\mathcal{E}_{S\leftarrow R}\big\{\hat{P}_{r}\big\}\big]\braket{\Psi_{r}\vert\hat{P}_{r_{q}}\vert\Psi_{r}}\nonumber \\
 & =\sum_{r}\textrm{Tr}\big[\hat{P}_{s}\mathcal{E}_{S\leftarrow R}\big\{\ket{\Psi_{r}}\braket{\Psi_{r}\vert\hat{P}_{r_{q}}\vert\Psi_{r}}\bra{\Psi_{r}}\big\}\big]\nonumber \\
 & =\textrm{Tr}\Big[\hat{P}_{s}\mathcal{E}_{S\leftarrow R}\Big\{\sum_{r}\hat{P}_{r}\hat{P}_{r_{q}}\hat{P}_{r}\Big\}\Big],
\end{align}
while the left-hand side of (\ref{eq:consistent}) is
\begin{equation}
p(s\vert r_{q})=\textrm{Tr}\big[\hat{P}_{s}\mathcal{E}_{S\leftarrow R}\big\{\hat{P}_{r_{q}}\big\}\big].
\end{equation}
We conclude that one set of maps that satisfies (\ref{eq:consistent})
are maps that incorporate a projective measurement along the $\{\hat{P}_{r}\}_{r}$
basis in their definition:
\begin{equation}
\mathcal{E}_{S\leftarrow R}\Big\{\sum_{r}\hat{P}_{r}\hat{P}_{r_{q}}\hat{P}_{r}\Big\}=\mathcal{E}_{S\leftarrow R}\big\{\hat{P}_{r_{q}}\big\}.
\end{equation}
Conceptually, this projective measurement ensures that the intermediate
composite state of the system and\emph{ }its environment re-factorize,
thus leading to Markov-like evolution.\footnote{Note that we could have instead inserted the projective measurement
step along the $\{\hat{P}_{r}\}_{r}$ basis into the map $\mathcal{E}_{R\leftarrow Q}$.
Either way, we demonstrate the existence of a set of maps satisfying
the consistency condition (\ref{eq:consistent}).} Putting all this together, we have
\begin{align*}
p(s\vert q) & \equiv\textrm{Tr}[\hat{P}_{s}\mathcal{E}_{S\leftarrow Q}\big\{\hat{P}_{q}\big\}]\\
 & =\textrm{Tr}[\hat{P}_{s}\mathcal{E}_{S\leftarrow R}\circ\mathcal{E}_{R\leftarrow Q}\big\{\hat{P}_{q}\big\}]\\
 & =\sum_{r_{q}}\textrm{Tr}[\hat{P}_{s}\mathcal{E}_{S\leftarrow R}\big\{\hat{P}_{r_{q}}\big\}]p(r_{q}\vert q)\\
 & =\sum_{r_{q}}p(s\vert r_{q})p(r_{q}\vert q)\\
 & =\sum_{r,r_{q}}p(s\vert r)\beta(r\vert r_{q})p(r_{q}\vert q),
\end{align*}
which with (\ref{eq:BornProbTransition}) yields
\begin{equation}
p(s\vert q)=\sum_{r}p(s\vert r)p(r\vert q).\label{eq:ChainProperty}
\end{equation}

The mutual information shared between the initial and final configurations
is
\begin{equation}
I(S:Q)=\sum_{s,q}p(s\vert q)p_{q}\log\left[\frac{p(s\vert q)}{p_{s}}\right].
\end{equation}
The mutual information between the initial and intermediate configurations
is
\begin{equation}
I(R:Q)=\sum_{r,q}p(r\vert q)p_{q}\log\left[\frac{p(r\vert q)}{p_{r}}\right].
\end{equation}
Using (\ref{eq:ChainProperty}), the difference between these two
quantities can be written as
\begin{align*}
I(S:Q)-I(R:Q) & =\sum_{s,q,r}p(s\vert r)p(r\vert q)p_{q}\left(\log\left[\frac{p(s\vert q)}{p_{s}}\right]-\log\left[\frac{p(r\vert q)}{p_{r}}\right]\right)\\
 & =\sum_{s,q,r}p(s\vert r)p(r\vert q)p_{q}\log\left[\frac{p(s\vert q)p_{r}}{p_{s}p(r\vert q)}\right].
\end{align*}
Using Jensen's inequality,\footnote{Jensen's inequality states that if $f(x)$ is a convex function of
its argument $x$, then the average of $f(x)$ provides an upper bound
for the original function applied to the average of its argument.
Here we apply Jensen's inequality to $-\log x$.} we have
\begin{align*}
I(S:Q)-I(R:Q) & \leq\log\left[\sum_{s,q,r}p(s\vert r)p(r\vert q)p_{q}\frac{p(s\vert q)p_{r}}{p_{s}p(r\vert q)}\right]\\
 & =\log\left[\sum_{s,q,r}p(s\vert r)p_{r}\frac{p(s\vert q)p_{q}}{p_{s}}\right]\\
 & =\log\left[\sum_{s}p_{s}\frac{p_{s}}{p_{s}}\right]\\
 & =0.
\end{align*}
We therefore arrive at a quantum version of the data-processing inequality,
\begin{equation}
I(S:Q)\leq I(R:Q),\label{eq:QDPI}
\end{equation}
capturing the idea that the information encoded in the system's initial
configuration is increasingly diluted as the system is ``processed.''

\subsection{A Holevo-Type Bound}

Let us recall the statement of Holevo's bound. Consider a quantum
system and let $X$ be a classical random variable with possible outcomes
$\{x\}_{x}$ and corresponding probability distribution $\{p_{x}\}_{x}$.
Suppose that $\{\hat{\rho}_{x}\}_{x}$ is a collection of density
matrices indexed by the possible outcomes $x$ of $X$, and let $\hat{\rho}$
be the correspondingly averaged density matrix:
\begin{equation}
\hat{\rho}\equiv\sum_{x}p_{x}\hat{\rho}_{x}.
\end{equation}
If we now measure a POVM $\{E_{Y}\}_{y}$ whose possible outcomes
$y$ form another classical random variable $Y$, then Holevo's bound
states that the classical mutual information between $X$ and $Y$
is bounded from above by the quantity
\begin{equation}
\chi\equiv S(\hat{\rho})-\sum_{x}p_{x}S(\hat{\rho}_{x}).
\end{equation}
That is,
\begin{equation}
I(X:Y)\leq\chi.\label{eq:HolevoBound}
\end{equation}

In the two-step process described in Section \ref{subsec:Quantum-Markovianity-and},
the mutual information between the initial configuration $\hat{\rho}_{Q}$
and the intermediate configuration $\hat{\rho}_{R}$ can be expressed
as
\begin{align}
I(R:Q) & =S(\hat{\rho}_{R})-J(R\vert Q)=S(\hat{\rho}_{R})-\sum_{q}p_{q}S\left(\hat{\rho}_{R\vert q}\right),\label{eq:HolevoChi}
\end{align}
where
\begin{equation}
\hat{\rho}_{R\vert q}=\sum_{r}\hat{P}_{r}\mathcal{E}_{R\leftarrow Q}\big\{\hat{P}_{q}\big\}\hat{P}_{r}.
\end{equation}
The quantity on the right-hand side of (\ref{eq:HolevoChi}) is clearly
an example of Holevo's $\chi$ quantity. We see that it emerges quite
naturally as an example of our newly defined mutual information, and
that Holevo's bound (\ref{eq:HolevoBound}) arises as a manifestation
of our quantum data-processing inequality (\ref{eq:QDPI}). The Holevo
bound's interpretation as a quantum version of the data-processing
inequality has been discussed before (see for example \cite{Hayden2004}).
Our dynamical interpretation of the bound provides another perspective
that avoids any explicit embedding of the system of interest into
a larger composite system. Instead, we capture the role of the broader
environment through the formalism of linear CPTP maps.

\section{Discussion\label{sec:Discussion}}

\subsection{Systems and Subsystems\label{subsec:Systems-and-Subsystems}}

Our focus in this paper has been on a dynamical interpretation of
quantum information in a system whose evolution is described by a
linear CPTP map. However, as mentioned in Section \ref{subsec:Quantum-Conditional-Probability},
the formalism is general enough to capture structural relationships
between composite quantum systems and their subsystems. To begin,
consider the parent system $AB$ formed from a pair of quantum subsystems
$A$ and $B$ and described by the density matrix
\begin{equation}
\hat{\rho}_{AB}=\sum_{m}p_{m}^{AB}\hat{P}_{m}^{AB},\qquad\hat{P}_{m}^{AB}=\ketbra{\Psi_{m}^{AB}}{\Psi_{m}^{AB}},
\end{equation}
where we include the parent system's label $AB$ on the system's eigenprojectors
$\hat{P}_{m}^{AB}$ and the corresponding probabilities $p_{m}^{AB}$.
The subsystem density matrices are related to $\hat{\rho}_{AB}$ via
the appropriate partial traces,
\begin{equation}
\hat{\rho}_{A}=\textrm{Tr}_{B}[\hat{\rho}_{AB}]=\sum_{a}p_{a}^{A}\hat{P}_{a}^{A},\qquad\hat{\rho}_{B}=\textrm{Tr}_{A}[\hat{\rho}_{AB}]=\sum_{b}p_{b}^{B}\hat{P}_{b}^{B},
\end{equation}
where the sets of eigenprojectors for subsystems $A$ and $B$ are
$\{\hat{P}_{a}^{A}\}_{a}$ and $\{\hat{P}_{b}^{B}\}_{b}$, respectively.

Quantum probabilities that conditionally link subsystem eigenstates
to a given eigenstate of the parent system are again defined using
(\ref{eq:QCP}), substituting the relevant partial trace for the linear
CPTP map in the formula. For instance, the conditional probability
that $\ket{\Psi_{a}^{A}}$ is the actual underlying state of subsystem
$A$ given that the underlying state of $AB$ is $\ket{\Psi_{m}^{AB}}$
is\footnote{We again adopt the language of the minimal modal interpretation, though
the mathematical content involves only textbook quantum theory.}
\begin{equation}
p(a\vert m)=\textrm{Tr}_{A}\big[\hat{P}_{a}^{A}\textrm{Tr}_{B}\big\{\hat{P}_{m}^{AB}\big\}\big].\label{eq:SubsystemQCP}
\end{equation}

As in Section \ref{subsec:Quantum-Conditional-Probability}, the partial
trace applied to system $AB$'s eigenprojector yields a density matrix
\begin{equation}
\hat{\rho}_{m}^{A}=\textrm{Tr}_{B}\big[\hat{P}_{m}^{AB}\big]=\sum_{a_{m}}p(a_{m}\vert m)\hat{P}_{a_{m}}^{A}.
\end{equation}
We have
\begin{equation}
\hat{\rho}_{A}=\sum_{m}p_{m}\hat{\rho}_{m}^{A}=\sum_{m}p_{m}\hat{\rho}_{A\vert m},\qquad\hat{\rho}_{A\vert m}=\sum_{a}\hat{P}_{a}^{A}\hat{\rho}_{m}^{A}\hat{P}_{a}^{A}.
\end{equation}
These relationships imply that the quantum entropy conditioned on
the parent state $\ket{\Psi_{m}^{AB}}$ satisfies the inequality
\begin{equation}
S(\hat{\rho}_{m}^{A})\leq J(A\vert m)=-\sum_{a}p(a\vert m)\log p(a\vert m)=S(\hat{\rho}_{A\vert m}),\label{eq:CondEntLessThan}
\end{equation}
due to the quantities $p(a\vert m)$ and $p(a_{m}\vert m)$ being
related via the doubly stochastic distribution
\begin{equation}
\beta(a\vert a_{m})=\vert\braket{\Psi_{a}^{A}\vert\Psi_{a_{m}}^{A}}\vert^{2}.
\end{equation}

It is interesting to examine the von Neumann entropy of $\hat{\rho}_{m}^{A}$,
\begin{equation}
S(\hat{\rho}_{m}^{A})=-\sum_{a_{m}}p(a_{m}\vert m)\log p(a_{m}\vert m),
\end{equation}
and to note that it is naturally interpreted as the entanglement entropy
of subsystem $A$ conditioned on the parent system $AB$ actually
occupying the pure state $\ket{\Psi_{m}^{AB}}.$ Note that when the
parent system is in a pure state, then $\hat{\rho}_{m}^{A}=\hat{\rho}_{A\vert m}$
and $J(A\vert m)$ is the entanglement entropy of subsystem $A$.

The full quantum conditional entropy is defined as
\begin{equation}
J(A\vert AB)=\sum_{m}p_{m}J(A\vert m).
\end{equation}
Therefore (\ref{eq:CondEntLessThan}) implies
\begin{equation}
\sum_{m}p_{m}S(\hat{\rho}_{m}^{A})\leq J(A\vert AB).
\end{equation}

There are also intriguing relationships between our quantum conditional
entropy (\ref{eq:QCPShannonInfo},\ref{eq:NewQCE}) and conditional
von Neumann entropy (\ref{eq:StandardQCE}). Observe that the inequality
(\ref{eq:MutualLessThanSinitial}) satisfied by our version of quantum
mutual information can be re-expressed as
\begin{equation}
S(\hat{\rho}_{A})-J(A\vert AB)\leq S(\hat{\rho}_{AB}),
\end{equation}
where the initial density matrix is taken to be $\hat{\rho}_{AB}$
and the final density matrix is $\hat{\rho}_{A}$. Rearranging terms
and applying the definition of conditional von Neumann entropy yields
\begin{equation}
-S(B\vert A)\leq J(A\vert AB).
\end{equation}
In the presence of entanglement, $S(B\vert A)$ may take on negative
values, leading to a \emph{positive }lower bound on $J(A\vert AB)$.
The result naturally captures the idea that when subsystems are entangled,
there is a non-zero minimal uncertainty about their states even given
information about the parent system.

\subsection{Generalizations of Quantum Conditional Probabilities}

Our definition of quantum conditional probability (\ref{eq:QCP})
involves the eigenprojectors of initial and final density matrices
(\ref{eq:InitialObjectiveDM}) and (\ref{eq:FinalObjectiveDM}), respectively.
However, as we described in Section \ref{sec:Background}, there are
infinitely many decompositions of a nontrivial density matrix. Thus,
we may consider quantities of the form
\begin{equation}
\mathscr{P}_{\mathcal{E}}(\rho\vert\kappa)=\textrm{Tr}\big[\hat{\Pi}_{\rho}^{R}\mathcal{E}_{R\leftarrow Q}\big\{\hat{\Pi}_{\kappa}^{Q}\big\}\big],\label{eq:GeneralizedQCP}
\end{equation}
where
\begin{equation}
\hat{\rho}_{Q}=\sum_{\kappa}\lambda_{\kappa}^{Q}\hat{\Pi}_{\kappa}^{Q},\qquad\hat{\rho}_{R}=\sum_{\rho}\lambda_{\rho}^{R}\hat{\Pi}_{\rho}^{R}
\end{equation}
are general convex decompositions of the system's initial and final
density matrices, respectively, with generic projection operators
\begin{equation}
\hat{\Pi}_{\kappa}^{Q}=\ketbra{\Phi_{\kappa}^{Q}}{\Phi_{\kappa}^{Q}},\qquad\hat{\Pi}_{\rho}^{R}=\ketbra{\Phi_{\rho}^{R}}{\Phi_{\rho}^{R}}.
\end{equation}
Note that such sets of projectors need not be orthogonal. However,
if we demand that the quantities (\ref{eq:GeneralizedQCP}) behave
as probabilities, then the set $\{\hat{\Pi}_{\rho}^{R}\}_{\rho}$
must resolve the identity:
\begin{equation}
\sum_{\rho}\hat{\Pi}_{\rho}^{R}=\mathbb{I}.
\end{equation}
Nevertheless, these quantities fail to act as fully satisfactory conditional
probabilities, as they do not obey a straightforward version of the
law of total probability. Instead we have
\begin{align}
\Lambda_{\rho}^{R} & \equiv\textrm{Tr}\big[\hat{\Pi}_{\rho}^{R}\hat{\rho}_{R}\big]\nonumber \\
 & =\textrm{Tr}\big[\hat{\Pi}_{\rho}^{R}\mathcal{E}_{R\leftarrow Q}\big\{\hat{\rho}_{Q}\big\}\big]\nonumber \\
 & =\sum_{\kappa}\textrm{Tr}\big[\hat{\Pi}_{\rho}^{R}\mathcal{E}_{R\leftarrow Q}\big\{\hat{\Pi}_{\kappa}^{Q}\big\}\big]\lambda_{\kappa}^{Q},
\end{align}
and thus
\begin{equation}
\Lambda_{\rho}^{R}=\sum_{\kappa}\mathscr{P}_{\mathcal{E}}(\rho\vert\kappa)\lambda_{\kappa}^{Q},
\end{equation}
where we generically have $\Lambda_{\rho}^{R}\neq\lambda_{\rho}^{R}$
due to the possible nonorthogonality of the projectors.

Despite their shortcomings as proper conditional probability distributions,
the quantities defined in (\ref{eq:GeneralizedQCP}) may yet be of
some interest for reasons we detail in Section \ref{sec:Conclusions}.

\section{Connections to Other Work\label{sec:Connections}}

\subsection{Relation to Causal Quantum Conditional States}

Interest in quantum analogues of information-theoretic quantities,
such as probabilities and entropies, dates back to the early work
of von Neumann \cite{vonNeumann1932}. Conditional counterparts of
these quantities have been studied in many works, often with the goal
of developing operators that capture quantum conditional expectations
\cite{Umegaki1962} or conditional versions of density matrices \cite{Cerf1997}.

Our quantum conditional probabilities (\ref{eq:QCP}) are most closely
related to a type of operator defined in \cite{Leifer2013} by Leifer
and Spekkens. By invoking the Choi-Jamio\l kowski isomorphism, Leifer
and Spekkens rephrase linear CPTP evolution in terms of what they
refer to as a ``causal quantum conditional state'' operator on a
double-copy of the system's Hilbert space. Our quantum conditional
probabilities turn out to be diagonal entries in their operator. We
explore these relationships in greater detail in \cite{Barandes2014}.

\subsection{Quantum Statistical Mechanics and Fluctuation Theorems\label{subsec:Quantum-Statistical-Mechanics}}

In \cite{Esposito2006}, Esposito and Mukamel investigate definitions
of work and heat, entropy production, and fluctuation theorems in
the context of open quantum systems. The authors' results rest on
their construction of quantum transition matrices that can be understood
in terms of the quantum conditional probabilities (\ref{eq:QCP})
used in this work. To see this connection, first we follow \cite{Esposito2006}
and describe the evolution of an open quantum system in terms of a
differential linear CPTP map $\mathcal{K}$ that defines the time
evolution of the system's density matrix,
\begin{equation}
\frac{d\hat{\rho}_{Q}}{dt}=\mathcal{K}\{\hat{\rho}_{Q}(t)\}.
\end{equation}
The quantum transition matrices of \cite{Esposito2006} can be expressed
as
\begin{equation}
w((q'\vert q);t)\equiv\textrm{Tr}\Big[\hat{P}_{q'}\mathcal{K}\{\hat{P}_{q}\}\Big].\label{eq:EspositoMukamelTransition}
\end{equation}
These transition rates satisfy a differential version of the law of
total probability,
\begin{equation}
\frac{dp_{q'}}{dt}=\sum_{q}w((q'\vert q);t)p_{q},
\end{equation}
which follows from the definition (\ref{eq:EspositoMukamelTransition})
and the relation
\begin{equation}
\frac{dp_{q}}{dt}=\textrm{Tr}\Big[\hat{P}_{q}\mathcal{K}\{\hat{\rho}_{Q}(t)\}\Big],
\end{equation}
together with the orthogonality of the operators $\hat{P}_{q}$ and
$d\hat{P}_{q}/dt$.

Using our definition (\ref{eq:QCP}) of quantum conditional probability,
we are formally able to reproduce the constructions of \cite{Esposito2006}
by considering a linear CPTP map $\mathcal{E}_{Q}^{t'\leftarrow t}$
that we interpret as evolving the density matrix $\hat{\rho}_{Q}(t)$
of a system $Q$ at time $t$ to the system's density matrix $\hat{\rho}_{Q}(t')$
at time $t'=t+\delta t$, for some small time interval $\delta t$.
The map $\mathcal{K}$ is then reproduced formally by taking
\begin{equation}
\mathcal{K}=\lim_{\delta t\to0}\frac{\mathcal{E}_{Q}^{t'\leftarrow t}-\textrm{Id}}{\delta t}.
\end{equation}
Similarly, the quantum transition matrix is given by
\begin{equation}
w((q'\vert q);t)=\lim_{\delta t\to0}\frac{p_{\mathcal{E}}(q_{t'}\vert q_{t})-\delta_{q'q}}{\delta t},
\end{equation}
where we have introduced indices $t'$ and $t$ indicating the explicit
time dependencies of the eigenprojectors of $\hat{\rho}_{Q}(t')$
and $\hat{\rho}_{Q}(t)$, respectively.

\subsection{Retrodiction in Quantum Theory}

In Section \ref{subsec:Quantum-Conditional-Probability}, we argued
that our quantum conditional probabilities (\ref{eq:QCP}) do not
generically satisfy Bayes' theorem due to the possible irreversibility
of the linear CPTP map $\mathcal{E}_{R\leftarrow Q}$ on which their
definition depends. The implications for retrodiction\textemdash inference
about past states given present conditions\textemdash are nuanced.
While $\mathcal{E}_{R\leftarrow Q}$ may not be reversible, there
may be situations in which a reverse evolution map can be defined,
as explored in the context of quantum fluctuation theorems by Aw,
Buscemi, and Scarini \cite{Aw2021}, which appeared while this work
was in preparation. Nonetheless, the generic asymmetry inherent in
the definition (\ref{eq:QCP}) typically precludes any retrodiction
based on our formulation of quantum conditional probabilities.\footnote{Watanabe raised questions about retrodiction even in situations where
Bayes' theorem is assumed to hold \cite{Watanabe2013}. Others have
attempted to address the generic time asymmetries in standard quantum
theory by formulating a retrodictive quantum theory. (See for example
\cite{Barnett2021} and references therein.) Our formulation, by contrast,
is built from standard elements of quantum theory, and thus time asymmetries
having to do with measurement processes or other open-system dynamics
are unavoidable.}

\section{Conclusions and Future Directions\label{sec:Conclusions}}

In this work, we utilized quantum conditional probabilities (\ref{eq:QCP})
that were first developed in \cite{Barandes2014} to define new forms
of quantum conditional entropy (\ref{eq:NewQCE}) and quantum mutual
information (\ref{eq:QCausalMutual}). We explored how these quantities
capture growth of entropy and loss of information as an open quantum
system evolves according to a linear CPTP evolution map.

Thanks to the existence of an underlying conditional probability distribution,
we were able to provide conceptually clear proofs of identities and
inequalities satisfied by our quantum conditional entropy and mutual
information, analogous to those satisfied by their classical counterparts.
By contrast, the traditional von Neumann conditional entropy and mutual
information generically lack any underlying conditional probabilities,
rendering their definitions and relationships conceptually unclear.

One limitation of our approach is that our quantum conditional probabilities
depend for their definition on the existence of a well-defined linear
CPTP map. For some of the results proved in this paper including (\ref{eq:RandomizingIncreasesEntropy}),
this limitation is benign because the claim itself is about a sub-class
of linear CPTP dynamics. For other proofs in this paper, like the
concavity of von Neumann entropy (\ref{eq:ConcavityvNE}), we were
able to introduce a linear CPTP map by hand without any loss of generality.

However, our derivation of the quantum data processing inequality
(\ref{eq:QDPI}) depended on the dynamics being described by a chain
of linear CPTP maps. The same is therefore true for our Holevo-type
bound in (\ref{eq:HolevoBound}), with $\chi$ given by the expression
on the right-hand side of (\ref{eq:HolevoChi}). In general, these
sorts of inequalities do appear to depend on the dynamics being at
least embeddable in some linear CPTP map \cite{Nielsen2000}. It would
be interesting to explore whether our approach could be used to study
more general forms of dynamics that can be systematically approximated
as analytically or numerically controllable deviations from linear
CPTP dynamics.

In light of the connections between our work and works such as \cite{Esposito2006},
as described in detail in Section \ref{subsec:Quantum-Statistical-Mechanics},
it would be interesting to explore the ways our quantum conditional
entropies and our other results, including our quantum data-processing
inequality, may be applied in understanding open-quantum system entropy
growth and fluctuation theorems.

Section \ref{subsec:Systems-and-Subsystems} explored intriguing connections
between our quantum conditional probabilities and standard quantum
information-theoretic concepts that arise from the rich structure
of system-subsystem relationships in quantum theory. In future work,
we will continue to explore these connections, along with related
concepts, such as quantum discord \cite{Ollivier2002}.

Despite their failure to reproduce the law of total probability, the
quantities (\ref{eq:GeneralizedQCP}) do satisfy the Kolmogorov axioms
for a basic probability distribution. They are also examples of more
general quantities of the form
\begin{equation}
F_{p}(\hat{A},\hat{B};\hat{K})=\textrm{Tr}\big[\hat{A}^{p}\hat{K}\hat{B}^{1-p}\hat{K}^{\dagger}\big],\label{eq:LiebQuantity}
\end{equation}
where $\hat{A}$ and $\hat{B}$ are positive semi-definite $N\times N$
matrices, $\hat{K}$ is a fixed $N\times N$ matrix, and $0\leq p\leq1$.
Lieb proved in \cite{Lieb1973} that trace quantities of the above
type are non-negative concave maps. Observe that when $\hat{A}$ and
$\hat{B}$ are taken to be projection operators with $p=1/2$, and
if $\hat{K}$ is one of the operators in a Kraus representation of
$\mathcal{E}_{R\leftarrow Q}$, then each term in the Kraus decomposition
of (\ref{eq:GeneralizedQCP}) is of the form (\ref{eq:LiebQuantity}).
Quantities such as (\ref{eq:LiebQuantity}) have been central to the
understanding of generalized entropies, particularly the properties
of quantum relative entropy, but their implications for the existence
of probability distributions in quantum theory seem worth exploring
further.

The properties of (\ref{eq:LiebQuantity}) provide one avenue for
proving the strong subadditivity of traditional von Neumann conditional
entropy. As a reminder to the reader, strong subadditivity is the
statement that the von Neumann conditional entropy of a system $Q$
given systems $R$ and $S$ is bounded from above by the von Neumann
conditional entropy of $Q$ given only $R$:
\begin{equation}
S(Q\vert RS)\leq S(Q\vert R).
\end{equation}
Strong subadditivity can then be used to prove many of the other properties
satisfied by quantum entropies and related quantities. Furthermore,
the surprising results of \cite{Almheiri2013} can also be seen as
a reflection of the strong subadditivity of von Neumann entropy. Given
these wide-ranging areas, we are quite interested in exploring whether
our quantum conditional probabilities and their associated quantum
conditional entropy can provide some new perspectives on strong subadditivity,
and hence shed some light on recent developments at the intersection
of quantum information and quantum gravity.

\section*{Acknowledgements}

We thank our departmental colleagues and staff for supporting our
work. D.K. thanks Darya Krym for useful discussions. Part of this
work was supported by the UMass Dartmouth Marine and Undersea Technology
Research Program (MUST) sponsored by the Office of Naval Research
(ONR) under grant N00014-22-1-2012. We would also like to thank our
anonymous reviewers for their insightful comments, which improved
our paper.

\appendix

\section*{Appendix: Proofs of Basic Information Inequalities\label{sec:Proofs-of-Basic}}

\subsection*{Properties of Doubly Stochastic Distributions}

A conditional probability distribution $p(y\vert x)$ is called doubly
stochastic if
\begin{equation}
\sum_{x}p(y\vert x)=1.
\end{equation}
If $p(y)$ and $p(x)$ are related via a doubly stochastic distribution,
\begin{equation}
p(y)=\sum_{x}p(y\vert x)p(x),
\end{equation}
then the Shannon entropy of $p(y)$ is greater than or equal to that
of $p(x)$. To see why, consider their difference:
\begin{equation}
H(X)-H(Y)=\sum_{x,y}p(y\vert x)p(x)\log\left(\frac{p(y)}{p(x)}\right).
\end{equation}
Using Jensen's inequality, we have
\begin{align}
H(X)-H(Y) & \leq\log\left(\sum_{x,y}p(y\vert x)p(x)\frac{p(y)}{p(x)}\right)\nonumber \\
 & =\log\left(\sum_{x,y}p(y\vert x)p(y)\right).
\end{align}
At this stage, we can use the double stochasticity of $p(y\vert x)$
to obtain
\begin{equation}
H(X)-H(Y)\leq\log\left(\sum_{y}p(y)\right)=0,
\end{equation}
and hence
\begin{equation}
H(X)\leq H(Y),
\end{equation}
as claimed.

While we have explicitly proved this result using classical notation,
the proof applies to von Neumann entropies linked via the quantum
conditional probabilities (\ref{eq:QCP}) defined in Section \ref{subsec:Quantum-Conditional-Probability}.

\subsection*{Non-Negativity}

The non-negativity of quantum conditional entropy follows directly
from its construction from non-negative conditional probabilities
that cannot be greater than one. Non-negativity of our form of quantum
mutual information arises by applying Jensen inequality to the definition
(\ref{eq:QCausalMutual}):
\begin{equation}
I_{\mathcal{E}}(R:Q)=-\sum_{q,r}p(r\vert q)p_{q}\log\left[\frac{p_{r}}{p(r\vert q)}\right]\geq-\log\left[\sum_{q,r}p(r\vert q)p_{q}\frac{p_{r}}{p(r\vert q)}\right]=-\log(1)=0.
\end{equation}
These arguments thus prove (\ref{eq:InformationPositivity}).

\subsection*{Linear CPTP Evolution Cannot Increase Mutual Information}

The difference between the quantum mutual information shared by the
initial and final configurations, on the one hand, and the von Neumann
entropy of the initial density matrix (\ref{eq:InitialObjectiveDM}),
on the other hand, is
\begin{align}
I_{\mathcal{E}}(R:Q)-S(\hat{\rho}_{Q}) & =-\sum_{q,r}p(r\vert q)p_{q}\log\left[\frac{p_{r}}{p(r\vert q)}\right]+\sum_{q}p_{q}\log p_{q}\nonumber \\
 & =\sum_{q,r}p(r\vert q)p_{q}\log\left[\frac{p(r\vert q)p_{q}}{p_{r}}\right].
\end{align}
The law of total probability (\ref{eq:TotalProbability}) gives us
\begin{equation}
p_{r}\geq p_{\mathcal{E}}(r\vert q)p_{q}.
\end{equation}
Thus, the monotonicity of the logarithm implies that
\begin{equation}
I_{\mathcal{E}}(R:Q)-S(\hat{\rho}_{Q})\leq\sum_{q,r}p(r\vert q)p_{q}\log\left[\frac{p_{r}}{p_{r}}\right]=0.
\end{equation}
We have thus proved (\ref{eq:MutualLessThanSinitial}).

\subsection*{Conditional Entropy Cannot Exceed Final Entropy}

The identity (\ref{eq:QCausalMutualAndQCE}) can be rewritten as
\begin{equation}
J_{\mathcal{E}}(R\vert Q)=S(\hat{\rho}_{R})-I_{\mathcal{E}}(R:Q).
\end{equation}
Due to the positivity of mutual information, we immediately have that
conditional entropy cannot exceed the final entropy of a system after
a linear CPTP process, (\ref{eq:ConditioningReducesEntropy}).

\bibliographystyle{unsrt}

\end{document}